\documentclass[%
 reprint,
 amsmath,amssymb,
 aps,
]{revtex4-2}

\usepackage{graphicx}% Include figure files
\usepackage{dcolumn}% Align table columns on decimal point
\usepackage{bm}% bold math
\usepackage{geometry}
\usepackage{xcolor}
\usepackage{mathtools}
\usepackage{amsmath}
\usepackage[normalem]{ulem}
\usepackage{amssymb}
\usepackage{babel}
\usepackage{tcolorbox}
\usepackage{comment}	
\usepackage{xr}
\usepackage{hyperref}% add hypertext capabilities
\newcommand{\hide}[1]{}
\usepackage{xcolor}
\definecolor{dark green}{rgb}{0.0, 0.5, 0.0}  % Define dark green color

\begin{document}

\preprint{APS/123-QED}

\title{Permutationally symmetric molecular aggregates}

\author{Sricharan Raghavan-Chitra}
\author{Arghadip Koner
}
\author{Joel Yuen-Zhou}%
 \email{joelyuen@ucsd.edu}
\affiliation{%
 Department of Chemistry and Biochemistry, University of California San Diego, La Jolla, California 92093
}%
\date{\today}% It is always \today, today,
             %  but any date may be explicitly specified

\begin{abstract}
\noindent Linear optical spectra of molecular aggregates are often approximated by classical optics methods such as the discrete-dipole approximation (DDA), coherent exciton scattering (CES), and coherent potential approximation (CPA), where the only quantum-mechanical input to the calculation is the linear susceptibility of the monomers. However, the limits of validity of these classical optics methods remain opaque. Here, starting from a quantum mechanical Hamiltonian for the aggregate, we identify a limit where DDA/CPA/CES is exact: all-to-all coupled permutationally symmetric aggregates of $N\to\infty$ monomers. The permutational symmetry of this molecular version of the Lipkin–Meshkov–Glick model, which is closely related to that of the molecular polariton problem of many identical molecules coupled to a single-cavity mode, allows us to borrow recent techniques developed for the latter. In particular, we identify a $1/N$ expansion that corrects the classical optics limit with finite $N$ corrections to the linear response of the aggregate. These corrections feature as Raman-like transitions of a single monomer. We illustrate these findings with calculations on the very physically-relevant setup of a homodimer. Our findings clarify how quantum optical features that go beyond classical optics can already be present in simple arrays of quantum emitters such as molecular aggregates. 
\end{abstract}

\maketitle

\section{\label{sec:level1}Introduction}

\noindent Molecular aggregates~\cite{jelley1936spectral, scheibe1937veranderlichkeit, wurthner2011j} are central to organic photophysics and optoelectronics, as intermolecular interactions can qualitatively reshape the optical response relative to that of isolated monomers. These collective effects govern processes such as exciton migration~\cite{wohlgemuth2020excitation, noriega2013general, patra2022vibronic}, charge separation~\cite{hill2000charge}, and radiative response~\cite{hestand2017molecular}, and thereby strongly influence the efficiency~\cite{xu2020precise}, stability~\cite{o2011ordered}, and emergent functionality~\cite{saikin2013photonics} of organic materials and devices~\cite{liu2010aggregate, bouachrine2025organic, ostroverkhova2016organic}. Their impact is evident across a wide range of applications, including OLEDs~\cite{mei2015aggregation}, organic solar cells~\cite{li2020molecular}, and organic photodetectors~\cite{spano2011vibronic}. Since the relevant photophysical processes are encoded most directly in optical spectra~\cite{bittner2022concerning, barford2017perspective, chen2009optical, noblet2021diagrammatic, reppert2020delocalization, devoe1965optical, spano1989nonlinear}, elucidating the structure--spectra relationship of molecular aggregates~\cite{spano2010spectral} remains a central objective. Such understanding is particularly important because molecular packing, intermolecular coupling, and excitonic architecture can profoundly modify spectral lineshapes and therefore provide a rational basis for the design of organic optoelectronic materials~\cite{ma2021organic}.\\

\noindent The theoretical description of aggregate spectra is challenging because the optical response generally reflects the interplay of electronic coupling, vibronic structure, and aggregate geometry. Foundational models of molecular aggregate spectroscopy~\cite{kasha1963energy, hestand2018expanded, clark2007role, anzola2019optical, dar2022theoretical, jansen2024electronically, kuhn2011quantum, hernandez2023modeling, ma2021organic} have yielded substantial insight under controlled approximations, often involving 
single-mode vibrations, harmonic potentials, and restricted
excitation manifolds. While these approaches successfully capture many essential mechanisms, more realistic settings may require treating multimode vibronic effects~\cite{hestand2018expanded}, anharmonic nuclear potentials~\cite{gohaud2005vibrational}, and large aggregate sizes on equal footing. In practice, this difficulty has motivated the widespread use of classical-optics approaches such as the discrete-dipole approximation (DDA)~\cite{devoe1964optical, devoe1965optical}, coherent exciton scattering (CES)~\cite{eisfeld2002j}, and coherent potential approximation (CPA)~\cite{soven1967coherent}. Notably, all these methods use the same analytical formula, in which the principal quantum-mechanical input is the linear susceptibility of the monomer. These methods have proved remarkably successful and, in favorable cases, even enable quantitatively predictive descriptions of aggregate spectra and excitation transport directly from monomer data~\cite{chenu2017construction}. Despite their utility, the domain of validity of these methods has remained insufficiently transparent. In particular, it is unclear when DDA, CPA, and CES become exact, what physical processes they neglect, and how corrections to them should be systematically organized from a microscopic quantum-mechanical theory.\\

\noindent In this work, we address these questions starting from a fully quantum-mechanical Hamiltonian for an all-to-all coupled, permutationally symmetric molecular aggregate. This model is a molecular analog of the Lipkin--Meshkov--Glick model and is closely related, through its symmetry structure, to the molecular polariton problem of many identical molecules coupled to a single cavity mode. This connection allows us to adapt recent techniques developed in the molecular polariton setting to the aggregate problem. In particular, we show that DDA, CPA, and CES emerge naturally as the $N\to\infty$ limit of the permutationally symmetric aggregate, thereby identifying a controlled limit in which these classical-optics descriptions become exact. This result clarifies the common microscopic origin of these seemingly distinct methods and rationalizes their success in modeling aggregate spectra. While this model may seem artificial for realistic aggregates in the presence of disorder, it constitutes a tractable theoretical construct that yields conceptually transparent limiting cases and admits an obvious physical realization in the ubiquitous case of molecular homodimers.\\

\noindent Beyond this classical-optics limit, we identify systematic corrections to the linear response of the aggregate, thereby elucidating the photophysical processes neglected at the classical-optics level. Remarkably, these corrections take the form of Raman-like transitions involving the vibrational structure of a single monomer, showing that linear spectra of finite aggregates can already encode quantum optical processes that lie beyond a purely classical description. We illustrate these corrections explicitly in the physically relevant case of a homodimer, where the Raman-mediated pathways provide the simplest nontrivial manifestation of the breakdown of the classical-optics limit. Our results therefore establish both the domain of validity of DDA/CPA/CES and the physical content of their leading corrections, clarifying how quantum optical effects beyond classical optics can already arise in simple molecular aggregates. \\

\noindent This article is structured as follows. In Sec.~\ref{sec: model}, we present the Hamiltonian of the all-to-all coupled aggregate. In Sec. ~\ref{sec: schwinger boson representation}, we reformulate this model in the Schwinger boson representation, which is enabled by the permutational invariance of the system; this representation dramatically facilitates the subsequent analysis. In Sec.~\ref{sec: exact linear response}, we compute the exact linear response for arbitrary \(N\) using a continued-fraction approach. In Sec.~\ref{sec: DDA/CPA/CES captures thermodynamic limit}, we analyze the linear spectra in the \(N\to\infty\) limit while keeping \(JN\) constant, and establish the connection of this limit to classical-optics methods such as DDA, CPA, and CES. Finally, in Sec.~\ref{sec: corrections to ces/cpa}, we investigate the corrections to these classical-optics descriptions in the physically relevant case of a homodimer.

\begin{figure}[ht!]
    \includegraphics[width=0.6\linewidth]{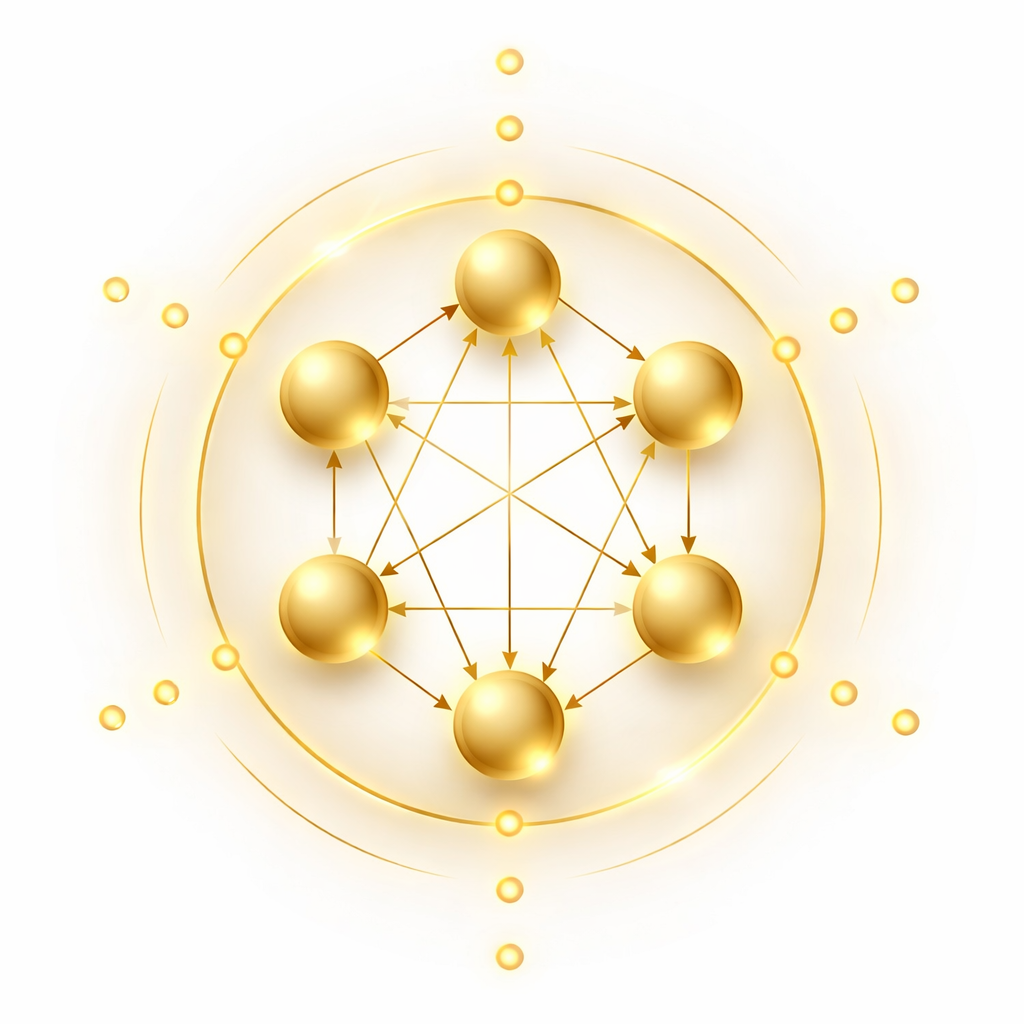}
    \caption{Schematic of an all-to-all coupled molecular aggregate. Each sphere denotes an identical monomer, while the bidirectional arrows indicate uniform, permutation-symmetric coupling between every pair of monomers, highlighting the collective and fully connected nature of the aggregate.}
\label{fig: all_to_all}
\end{figure}

\section{Model} 
\label{sec: model}

\begin{figure*}[ht!]
    \includegraphics[width=1\linewidth]{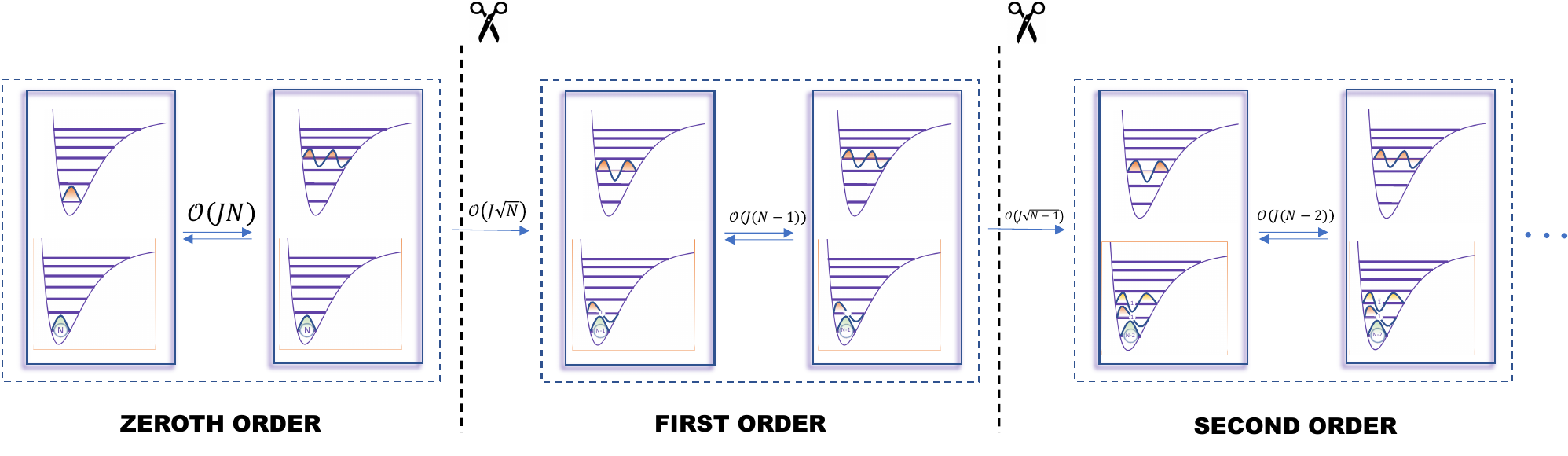}
    \caption{ Hierarchy of dynamical timescales in an all-to-all coupled molecular aggregate. Starting from a permutationally symmetric initial condition with one electronic excitation delocalized over $N$ monomers, exciton exchange processes that preserve the number of molecules with ground-state vibrational excitations occur at an enhanced collective rate scaling as $\mathcal{O}(JN)$. In contrast, pathways that involve the creation or annihilation of molecules with ground-state phonons are parametrically slower, with characteristic amplitudes scaling as $\mathcal{O}(J\sqrt{N})$. This separation of scales naturally motivates a systematic hierarchy of approximations—ranging from zeroth order to higher orders—in which progressively higher-order processes, involving the creation of multiple molecules with ground-state phonons, are successively neglected, as schematically illustrated in the figure. Note that the zeroth, first, and second order dynamics utilize one, two, and three particle states of the Philpott basis~\cite{philpott1971theory} (a widely used basis in the molecular aggregates community), respectively}
\label{fig: cut-e diagrams}
\end{figure*}

\noindent We begin by considering an arbitrary ensemble of identical, all-to-all coupled molecular aggregates (schematically shown in Fig.~\ref{fig: all_to_all}) , which at a formal level may be regarded as a generalized extension of the Lipkin–Meshkov–Glick model~\cite{dusuel2005continuous, dusuel2004finite} in the presence of vibronic coupling. The Hamiltonian of this system can be written as
\begin{equation}
\label{eq: overall_hamiltonian_first_quantized_language}
\hat{H} = \sum_{i=1}^{N+1} \hat{H}_m^{(i)} + \sum_{i \neq j}^{N+1} \hat{H}_I^{(ij)} .
\end{equation}
Here, $\hat{H}_m^{(i)}$ denotes the molecular Hamiltonian of the $i$th molecule, while $\hat{H}_I^{(ij)}$ represents the interaction between the $i$th and $j$th molecules. For analytical convenience, we consider a system composed of $N+1$ monomers rather than the conventional $N$; this choice avoids the appearance of cumbersome numerical prefactors, as will become evident later in Fig.~\ref{fig: cut-e diagrams}. \\

\noindent Now, under the Born--Oppenheimer approximation, the molecular Hamiltonian of the $i$th molecule is given by
\begin{equation}
\begin{aligned}
\hat{H}_m^{(i)} 
= \hat{T}_i 
+ V_g(q_i)\lvert g_i\rangle\langle g_i\rvert 
+ V_e(q_i)\lvert e_i\rangle \langle e_i\rvert .
\end{aligned}
\end{equation}

\noindent
Here, $\hat{T}$ is the nuclear kinetic energy operator, $V_{g/e}$ denote the ground and excited potential energy surfaces (PESs), and $q_i$ represents the set of all intramolecular vibrational coordinates associated with the $i$th molecule. Furthermore, the interaction Hamiltonian captures excitonic coupling between the $i$th and $j$th molecules~\cite{spano2011vibronic}:
\begin{equation}
\begin{aligned}
\hat{H}_I^{(ij)} = J \lvert e_i\rangle\langle e_j\rvert ,
\end{aligned}
\end{equation}

\noindent
where $J$ denotes the interaction coupling strength, assumed to be identical for any pair of monomers.\\

\section{Schwinger boson representation}
\label{sec: schwinger boson representation}

\noindent The first-quantized many-body wavefunction of this system may be expanded in a complete tensor-product basis constructed from single-molecule states $\varphi_\nu$ associated with each molecule in the ensemble. Explicitly, the time-dependent many-body wavefunction takes the form
\begin{equation}
\Psi(\vec{q},t)
=
\sum_{\nu_{1}\nu_{2}\cdots\nu_{N+1}}
A_{\nu_{1}\nu_{2}\cdots\nu_{N+1}}(t)\,
\prod_{k=1}^{N+1}
\varphi_{\nu_{k}}(q_{k}).
\end{equation}\\

\noindent  For an initial wavefunction that is permutationally symmetric, indeed, the only physically relevant class in the context of far-field spectroscopy, this symmetry is conserved under the unitary time evolution generated by $\hat{H}$. Consequently, the many-body wavefunction remains symmetric under the exchange of any pair of molecular coordinates, i.e.,
\begin{equation}
\Psi( \ldots, q_k, \ldots, q_l, \ldots, t)
=
\Psi( \ldots, q_l, \ldots, q_k, \ldots, t).
\end{equation}

\noindent This observation implies that the system dynamics is restricted to the permutationally symmetric sector of the full Hilbert space. In terms of the bosonic symmetrization operator, the symmetric many-body wavefunction can be written as
\begin{equation}
\Psi_{+}(\vec{q},t)
=
\sum_{\nu_{1}\nu_{2}\cdots\nu_{N+1}}
\tilde{A}_{\nu_{1}\nu_{2}\cdots\nu_{N+1}}(t)
\hat{S}_{+}
\prod_{k=1}^{N+1}
\varphi_{\nu_{k}}(q_{k}),
\end{equation}
where $\hat{S}_{+}$ is defined by the permanent over molecule permutations:

\begin{equation}
\hat{S}_{+}\prod_{k=1}^{N+1}\varphi_{\nu_k}(q_k)
=
\left|
\begin{array}{cccc}
\varphi_{\nu_1}(q_1) & \varphi_{\nu_1}(q_2) & \cdots & \varphi_{\nu_1}(q_{N+1}) \\
\varphi_{\nu_2}(q_1) & \varphi_{\nu_2}(q_2) & \cdots & \varphi_{\nu_2}(q_{N+1}) \\
\vdots & \vdots & \ddots & \vdots \\
\varphi_{\nu_{N+1}}(q_1) & \varphi_{\nu_{N+1}}(q_2) & \cdots & \varphi_{\nu_{N+1}}(q_{N+1})
\end{array}
\right|_{+}.
\end{equation}\\

\noindent At this stage, it becomes evident that the second-quantized formalism provides the natural framework for describing the system, analogous to its use for bosons. This formulation circumvents the explicit symmetrization and normalization procedures required in the first-quantized treatment of many-body wavefunctions. Such complications arise from attempting to assign a time-dependent state to each individual molecule. However, because the molecules are indistinguishable by construction, the physically meaningful description is instead given in terms of occupation numbers specifying how many molecules populate each single-molecule state.\\

\noindent In the second-quantized representation, this information is encoded in occupation-number states of the form $| n_{\nu_1}, n_{\nu_2}, n_{\nu_3}, \ldots \rangle$, where $n_{\nu_i}$ denotes the number of molecules occupying the single-molecule state $\varphi_{\nu_i}$. Operators are mapped to this representation through bosonic annihilation and creation operators $\hat{\beta}_{\nu_i}$ and $\hat{\beta}_{\nu_i}^\dagger$. In the first-quantized framework, the corresponding operators insert or remove a single-molecule function $\varphi_{\nu_i}$ from the symmetrized many-body wavefunction while preserving overall bosonic symmetry. In contrast, within the second-quantized formalism, $\hat{\beta}_{\nu_i}$ and $\hat{\beta}_{\nu_i}^\dagger$ lower and raise the occupation number of the state $\varphi_{\nu_i}$ by one, respectively, thereby annihilating or creating a molecule in that single-molecule state. \\

\noindent Furthermore, any symmetric sum of single-molecule operators,
$\hat{O} = \sum_{k=1}^{N+1} \hat{o}^{(k)}$,
admits a systematic representation in the second-quantized formalism. Specifically, employing bosonic creation and annihilation operators, one obtains
\begin{equation}
\hat{O}
\;\longrightarrow\;
\sum_{\nu_i \nu_j}
\langle \nu_i | \hat{o} | \nu_j \rangle
\hat{\beta}_{\nu_i}^{\dagger}\hat{\beta}_{\nu_j}.
\end{equation}

\noindent In an analogous manner, two-molecule operators are expressed as
\begin{equation}
\sum_{k \neq l}^{N+1} \hat{o}_2^{(kl)}
\;\longrightarrow\;
\sum_{\nu_a \nu_b \nu_c \nu_d}
\langle \nu_a, \nu_b | o_2 | \nu_c, \nu_d \rangle
\hat{\beta}_{\nu_a}^{\dagger}
\hat{\beta}_{\nu_b}^{\dagger}
\hat{\beta}_{\nu_c}
\hat{\beta}_{\nu_d}.
\end{equation}

\noindent Adopting the vibronic eigenstates $|\varphi_i^{(g)}, g\rangle$ and $|\varphi_i^{(e)}, e\rangle$ as the single-molecule basis, and invoking the mappings defined above, we obtain the reformulated Hamiltonian of the all-to-all coupled molecular aggregate in terms of Schwinger bosons,
\begin{equation}
\begin{aligned}
H &= \sum_{n=0}^{M}\omega_{g,n}\,b_{n}^{\dagger}b_{n}
   + \sum_{m=0}^{M}\omega_{e,m}\,B_{m}^{\dagger}B_{m} \\
&\quad + J \sum_{nm\tilde{n}\tilde{m}}^{M}
\langle\varphi_{n}^{(g)} \mid \varphi_{\tilde{m}}^{(e)}\rangle
\langle\varphi_{m}^{(e)} \mid \varphi_{\tilde{n}}^{(g)}\rangle\,
b_{n}^{\dagger} B_{m}^{\dagger} B_{\tilde{m}} b_{\tilde{n}} .
\end{aligned}
\label{eq: bosonic_all_to_all}
\end{equation}

\noindent Here, $|\varphi_i^{(g)}\rangle$ and $|\varphi_i^{(e)}\rangle$ denote the $i$-th vibrational adiabatic eigenstates of the molecular Hamiltonian within the electronic ground and excited manifolds, respectively. The operators $\hat{b}_i$ and $\hat{B}_j$ annihilate a molecule in the ground- and excited-state vibronic adiabatic states $|\varphi_i^{(g)}, g\rangle$ and $|\varphi_j^{(e)}, e\rangle$, respectively. Here, $M$ denotes the number of vibrational basis states, $\omega_{g/e,n}$ are the corresponding vibronic eigenenergies, and $\langle \varphi_n^{(g)} | \varphi_{\tilde{m}}^{(e)} \rangle$ are the Franck--Condon factors associated with the relevant optical transitions. The interaction term correspondingly assumes the form of a four-body bosonic operator. These bosonic operators obey the canonical commutation relations $[ b_{n_1}, b_{n_2}^\dagger ] = \delta_{n_1,n_2}$, $[ B_{m_1}, B_{m_2}^\dagger ] = \delta_{m_1,m_2}$, and $[ b_{n_1}, B_{m_2}^\dagger ] = [ B_{m_1}, b_{n_2}^\dagger ] = 0$.\\

\noindent This mapping may be viewed as a natural extension of the Schwinger boson representation for spin systems to molecular systems with more than two internal states. It has previously been employed in the description of molecular polariton systems by several authors~\cite{perez2025radiative, gegg2016efficient, shammah2018open, zeb2022efficient, silva2022permutational, pizzi2023light, sukharnikov2023second}. Within this notation, the many-body basis states $| n_0 n_1 \cdots n_M,\; n'_0 n'_1 \cdots n'_M \rangle$ are eigenstates of the noninteracting Hamiltonian, i.e., in the limit $J = 0$. Here, $n_i$ and $n'_i$ denote the number of molecules occupying the vibronic states $| \varphi_i^{(g)}, g \rangle$ and $| \varphi_i^{(e)}, e \rangle$, respectively. These basis states do not encode the evolution of individual molecules; rather, they are fully characterized by the occupation numbers associated with the accessible vibronic configurations.\\

\noindent It can be readily verified that the permutationally symmetric molecular aggregate Hamiltonian conserves both the total number of molecules and the number of excitations. This follows from the operators
$ \hat{N}_{\mathrm{mol}}=\sum_{i}^{M}\left(\hat{b}_i^\dagger \hat{b}_i+\hat{B}_i^\dagger \hat{B}_i\right), 
\qquad
\hat{N}_{\mathrm{exc}}=\sum_{i}^{M}\hat{B}_i^\dagger \hat{B}_i+\hat{a}^\dagger \hat{a},$
which satisfy
$[\hat{N}_{\mathrm{mol}},\hat{H}]=[\hat{N}_{\mathrm{exc}},\hat{H}]=0 .$\\

\section{Exact linear response}
\label{sec: exact linear response}

\noindent The linear optical response of the all-to-all–coupled aggregate can be computed from the following expression (Ref.~X):

\begin{equation}
    \sigma(\omega)\propto-\Im\langle\hat{\mu}\hat{G}(\omega)\hat{\mu}\rangle.
\end{equation}

\noindent Here, the expectation value is taken with respect to the initial ground state of the aggregate,
$\left|N+1,0,\dots,0;\,0,\dots,0\right\rangle$, expressed in the bosonic occupation-number representation. The operator $\hat{\mu}$ denotes the collective transition dipole moment that couples to the incident electromagnetic field, and $\hat{G}(\omega) = (\omega - \hat{H} + i\epsilon)^{-1}$ is the retarded Green’s function of the aggregate.\\

\noindent Because the Hamiltonian in Eq.~\ref{eq: bosonic_all_to_all} conserves the total number of electronic excitations, the linear optical response of the all-to-all coupled aggregate, initialized in its zero-temperature ground state, is restricted to the single-excitation manifold. Consequently, the linear response is fully determined by the Hamiltonian projected onto the first electronically excited manifold. In the occupation-number basis $| n_0 n_1 \cdots n_M,\; n'_0 n'_1 \cdots n'_M \rangle$, the Hamiltonian projected onto this manifold acquires a block-tridiagonal structure:

% \dgreen{Show that the Hamitlonian is block diagonal}

% \dgreen{Explain the couplings present in Fig. 2.}

\begin{equation}
\hat{H}^{(1)} \equiv
\begin{pmatrix}
H_{e,0} & v_{0} & 0 & \cdots & 0 \\
v_{0}^{\dagger} & H_{e,1} & v_{1} & \ddots & \vdots \\
0 & v_{1}^{\dagger} & H_{e,2} & \ddots & 0 \\
\vdots & \ddots & \ddots & \ddots & v_{N-1} \\
0 & \cdots & 0 & v_{N-1}^{\dagger} & H_{e,N}
\end{pmatrix}.
\label{eq: block tri-diagonal matrix}
\end{equation}\\

\noindent Here, the diagonal blocks are denoted by $H_{e,k}$, while the off-diagonal blocks are denoted by $v_k$. The basis states spanning the subspace associated with $H_{e,k}$ consist of configurations containing a single electronically excited molecule (i.e., $\sum_{i=0}^{M} n'_i = 1$) and $k$ molecules with vibrational excitations (i.e., $\sum_{i=1}^{M} n_i = k$). We denote these basis states by $\{ |{\alpha_k}\rangle \}$; in the language of the Philpott basis~\cite{philpott1971theory}, which is frequently used in molecular aggregate studies~\cite{spano2002absorption}, $\{|\alpha_{k}\rangle\}$ corresponds to the (k+1)-particle basis (one exciton and k molecules with ground state vibrations). Notably, $H_{e,k}$ is not diagonal in this basis.\\

\noindent Using the matrix representation in Eq.~\ref{eq: block tri-diagonal matrix}, we now elucidate the underlying photophysical processes contributing to the linear absorption spectrum. Optical excitation prepares the system in one-particle states of the form $|{N,0,\dots,0;\dots,1_m,\dots}\rangle$, which belong to the $H_{e,0}$ manifold. Consequently, as we discuss below, the short-time response is governed by the off-diagonal couplings within $H_{e,0}$, whereas the long-time dynamics is governed by $v_k$ couplings.\\

\noindent This hierarchy of timescales is illustrated schematically in Fig.~\ref{fig: cut-e diagrams}, which represents Eq.~\ref{eq: block tri-diagonal matrix} in terms of shifted harmonic oscillator (SHO) diagrams. Each of these diagrams corresponds to a basis state $\{ |{\alpha_k}\rangle \}$ of the aggregate Hamiltonian. Owing to the separation of timescales discussed below, we classify the schematic blocks as zeroth\nobreakdash-, first\nobreakdash-, and second-order, etc., reflecting a controlled hierarchy of approximations.\\

\noindent At zeroth order, the SHO diagram represents the manifold $\{ |{\alpha_{k=0}}\rangle \}$, consisting of configurations of the one-particle states, in which one molecule is electronically excited while all others remain in the global ground state. Within this manifold, the interaction term in Eq.~\ref{eq: block tri-diagonal matrix} couples any pair of basis states (including diagonal self-couplings) with an effective strength scaling as $\mathcal{O}(JN)$, as indicated by the arrows in the zeroth-order panel of Fig.~2. A pedagogical example of this scaling is provided in the supplementary information section IIA.\\

\noindent At first order, the relevant SHO diagram corresponds to the permutationally symmetric manifold $\{ |{\alpha_{k=1}}\rangle \}$, comprising two-particle states in which one molecule is electronically excited and a second molecule carries a vibrational excitation in the electronic ground state. In this sector, the interaction couples basis states with a strength scaling as $\mathcal{O}\!\big(J(N-1)\big)$, as depicted in the first-order panel of Fig.~2. A pedagogical example of this scaling is provided in the supplementary information section IIA. Further, this reasoning extends straightforwardly to higher-order manifolds.\\

\noindent In addition to the intra-manifold couplings discussed above, Eq.~\ref{eq: block tri-diagonal matrix} contains weaker off-diagonal contributions represented by the matrices $v_k$. These terms couple basis states belonging to different vibrational manifolds, i.e., states that differ in the number of molecules carrying vibrational excitations. Parametrically, these inter-manifold couplings are weaker than the corresponding intra-manifold couplings.\\

\noindent For example, the coupling strength between states within the manifold $\{ |{\alpha_{k=0}}\rangle \}$ scales as $\mathcal{O}(JN)$. In contrast, the coupling between a state in $\{ |\alpha_{k=0}\rangle \}$ and a state in $\{ \alpha_{k=1}\rangle \}$, where, in addition to one electronically excited molecule, another molecule carries a vibrational excitation, scales as $\mathcal{O}(J\sqrt{N})$. This reduced coupling strength is represented by the arrows connecting the zeroth- and first-order diagrams in Fig.~2. Similarly, the matrix $v_2$ contains terms that scales as $\mathcal{O}(J\sqrt{N-1})$, depicted by the arrows linking the first- and second-order diagrams. The same structural hierarchy and scaling behavior extend systematically to higher-order manifolds.\\

\noindent Hence, Fig.~2 schematically illustrates the block-structured form of Eq.~\ref{eq: block tri-diagonal matrix}: for $k \ll N$, the dominant couplings are those within each manifold, encoded in $H_{e,k}$, whereas the inter-manifold couplings contained in $v_k$ are parametrically weaker.\\

\noindent This hierarchy of timescales is analogous to that encountered in polaritonic systems, which underpin the Collective Dynamics Using Truncated Equations (CUT-E) method~\cite{perez2023simulating}. The linear response of systems exhibiting such a separation of timescales naturally admits a continued-fraction representation~\cite{koner2025hidden}. This representation enables efficient evaluation of the optical response, even for aggregates comprising an arbitrarily large number of monomers coupled in an all-to-all configuration (see supplementary information section I):

\begin{widetext}
\begin{equation}
\sigma(\omega)\propto - \Im \Bigg\langle \mu \frac{1}{\omega-H_{e,0}+i\frac{\gamma}{2}-v_{0}\frac{1}{\omega-H_{e,1}+i\frac{\gamma+\gamma_{v}}{2}-v_{1}\frac{1}{\frac{\ddots}{\omega-H_{e,N-1}+i\frac{\gamma+\gamma_{v}}{2}-v_{N-1}\frac{1}{\omega-H_{e,N}+i\frac{\gamma+\gamma_{v}}{2}}v^\dagger_{N-1}}}v_{1}^{\dagger}}v_{0}^{\dagger}}\mu\Bigg\rangle,
\label{eq: exact_all_to_all_response}
\end{equation}
\end{widetext}

\noindent where $\gamma$ and $\gamma_v$ are the homogeneous decay rates of the electronic and vibrational excitations in the system. Furthermore, this continued-fraction representation naturally enables a systematic $1/N$ expansion, closely paralleling analogous developments in light–matter systems~\cite{perez2025cut, koner2025hidden, schellenberger2026infinity} and broader condensed-matter settings~\cite{barberena2025generalized}.

\section{DDA/CPA/CES captures thermodynamic limit}
\label{sec: DDA/CPA/CES captures thermodynamic limit}

\begin{figure*}[ht!]
    \includegraphics[width=1\linewidth]{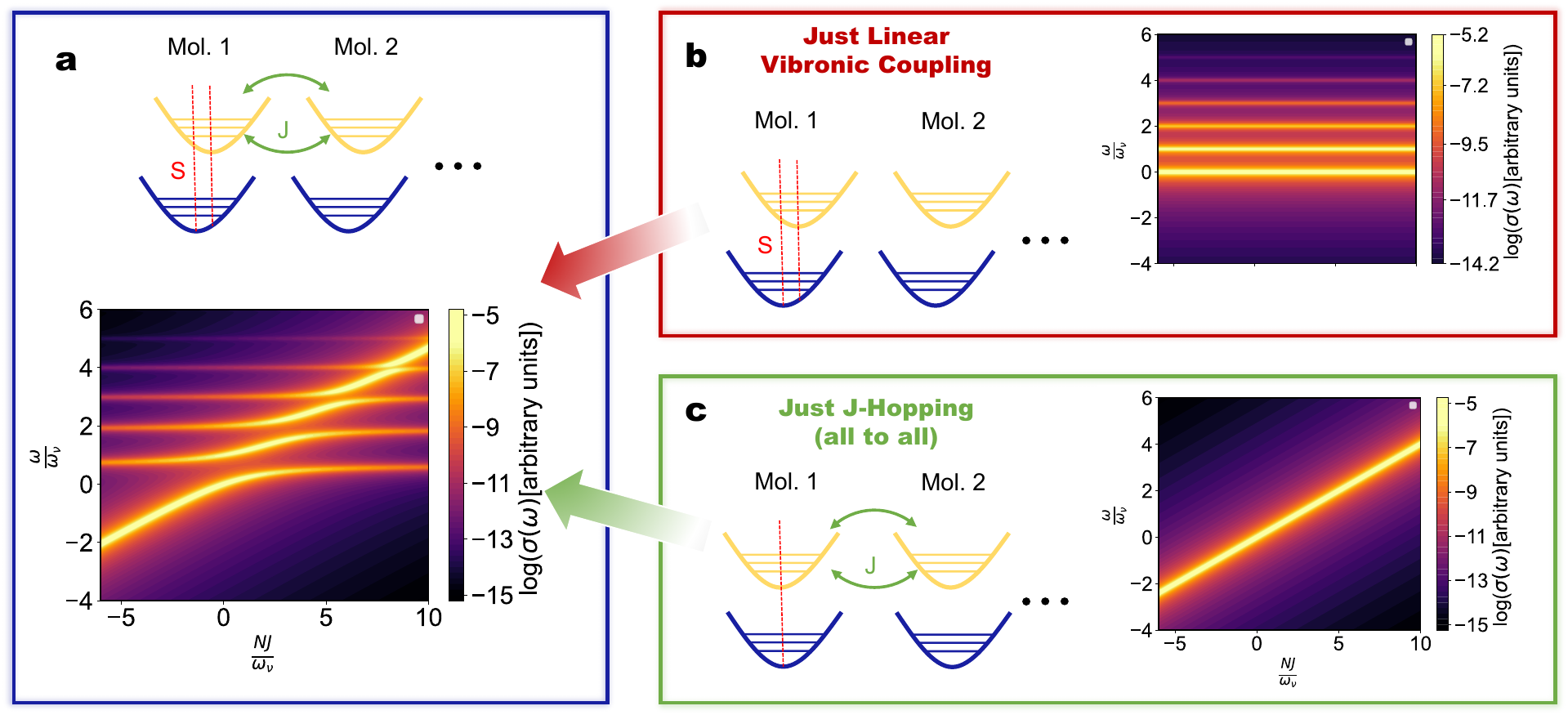}
    \caption{Linear absorption spectra of all-to-all coupled molecular aggregates are shown on a logarithmic intensity scale across distinct coupling regimes, with accompanying schematics that elucidate the underlying interactions. Panel (a) corresponds to the full regime in which both vibronic coupling and inter-monomer electronic coupling $J$ are simultaneously active. The characteristic level splittings and spectral features observed in this regime can be rationalized by comparison with the limiting cases shown in panels (b) and (c), where only vibronic coupling or only electronic $J$ coupling is retained, respectively. Together, these panels elucidate how the combined action of vibronic and electronic interactions shapes the aggregate absorption profile. Here, we have considered the Huang-Rhys factor $S=1/2$.}
\label{fig: linear_spectra_all_to_all}
\end{figure*}

\noindent In this section, we analyze the linear absorption spectrum of the all-to-all--coupled aggregate in the many-monomer limit, $N \to \infty$, at fixed collective coupling strength $JN$, and establish its connection to DDA/CPA/CES.\\

\noindent In this limit, Eq.~\ref{eq: exact_all_to_all_response} simplifies: as $N \to \infty$ with $JN$ held fixed, the ratio of intra-manifold to inter-manifold couplings diverges. Consequently, processes mediated by $v_1$ do not contribute to the linear response. We refer to this as the zeroth-order approximation, wherein only Rayleigh-type processes involving one-particle states occur and electronic excitations are exchanged elastically, without generating vibrational quanta in the electronic ground state. Higher-order terms incorporate slower Raman-type processes (Fig.~\ref{fig: cut-e diagrams}). Under the zeroth-order approximation, the response is given by

\begin{equation}
    \begin{aligned}
   \sigma_{\text{zeroth}}(\omega)&\propto-\Im \Big \langle \mu\frac{1}{\omega-H_{e,0}+i\frac{\gamma}{2}} \mu\Big \rangle.
    \end{aligned}
    \label{eq: zeroth_order_spectra}
\end{equation}

\noindent This absorption spectrum (Fig.~\ref{fig: linear_spectra_all_to_all}(a)) encodes both the vibronic structure of the electronically excited monomer and the collective aggregate response, manifested as a shift of the dominant absorption peak whose direction depends on the sign of $J$. The interplay between these contributions gives rise to characteristic level crossings~\cite{polyutov2012exciton,butkus2014vibronic,wang2014simple,schroeter2015unraveling} as the dimensionless ratio 
$\frac{NJ}{\omega_v}$ is varied. By examining the limiting regimes of purely electronic coupling (Fig.~\ref{fig: linear_spectra_all_to_all}(b)) and purely vibronic coupling (Fig.~\ref{fig: linear_spectra_all_to_all}(c)), we isolate the physical mechanisms underlying these crossings. Notably, these features are already apparent in the analytical expression obtained from Eq.~\ref{eq: zeroth_order_spectra}, given by

\begin{equation}
\begin{aligned}
\sigma_{\text{zeroth}}(\omega)
&\propto -\Im \langle \hat{\mu} \hat{G}_{\text{zeroth}}(\omega)\hat{\mu}  \rangle \\
&\propto -\Im \frac{\left\langle \hat{\mu} \hat{g}(\omega) \hat{\mu} \right\rangle}
{1 - NJ \left\langle \hat{\mu} \hat{g}(\omega) \hat{\mu} \right\rangle}.
\end{aligned}
\label{eq: DDA/CPA/CES formula}
\end{equation}

\noindent Here, $g(\omega)$ denote the Green’s function of the $N$ uncoupled monomers (supplementary information section III), respectively. As a result, the aggregate spectrum admits an explicit decomposition into intrinsic monomer absorption characteristics and coupling-induced spectral modifications. This compact expression appears in the literature on aggregates~\cite{briggs1970sum, briggs1971band, devoe1964optical, devoe1965optical} and disordered alloys~\cite{soven1967coherent, velicky1969theory, soven1969contribution}, commonly referred to as the DDA/CPA/CES approximation, as well as the classical optical limit of polaritons~\cite{yuen2024linear}. In those contexts, similar formulas to Eq.~\ref{eq: DDA/CPA/CES formula} are obtained via mathematical approximations and have been employed to compute multichromophoric Förster resonance energy transfer rates with remarkable efficiency~\cite{chenu2017construction}. Beyond computational advantages, Chenu et al. demonstrated that these approximations show excellent agreement with results obtained from stochastic path-integral formalisms. Moreover, such approximations have been applied to model molecular aggregates of arbitrary size and intermolecular connectivity~\cite{kato1998theoretical, eisfeld2006j, eisfeld2002j}. By contrast, our analysis reveals that, for all-to-all coupled aggregates in the thermodynamic limit, this reduction is not merely approximate but becomes exact. This perspective further clarifies the parameter regimes explored by Chenu et al. (fig. 2 in ref.~\cite{chenu2017construction}), which effectively realize Förster-type interactions within an all-to-all coupled aggregate and therefore naturally yield excellent agreement with exact simulations. \\

\noindent Beyond unifying DDA, CPA and CES and clarifying the origin of CPA’s success in the aggregate literature, the thermodynamic-limit all-to-all model furnishes a physically transparent and exactly solvable surrogate for the CPA based optical response (supplementary information section IIIA). This surrogate provides a principled framework for interpreting the approximate dynamics of molecular aggregates with arbitrary size and intermolecular connectivity in terms of the underlying collective excitation pathways of an all-to-all system.\\

\noindent In the following section, we analyze the approximations underlying the DDA/CPA/CES frameworks and elucidate why the linear absorption spectra obtained within these approaches exactly reproduce those of all-to-all coupled aggregates in the thermodynamic limit. We further show how systematic corrections to these approximations can be constructed, enabling the accurate modeling of more realistic aggregate systems.\\

\section{corrections to DDA/CPA/CES}
\label{sec: corrections to ces/cpa}

\noindent The relationship between aggregate and monomer spectra obtained within the DDA/CPA/CES framework originates from a mean-field–type approximation~\cite{briggs1970sum, eisfeld2002j, chenu2017construction}, as formulated in the original DDA/CPA/CES literature. From a physical standpoint, this mathematical approximation amounts to neglecting specific contributions to the molecular Green’s function, namely, those associated with optical processes that generate ground-state vibrational excitations. As a consequence, the spectra computed within DDA/CPA/CES coincide with those of an all-to-all coupled aggregate in the thermodynamic limit, where such processes exactly vanish.\\

\noindent Importantly, reinstating these neglected contributions naturally yields systematic corrections beyond the DDA/CPA/CES, a direction that has not been explored previously. In this section, we explicitly construct and analyze these corrections in the simplest nontrivial setting: a molecular dimer.\\

\begin{figure}[ht!]
    \includegraphics[width=1\linewidth]{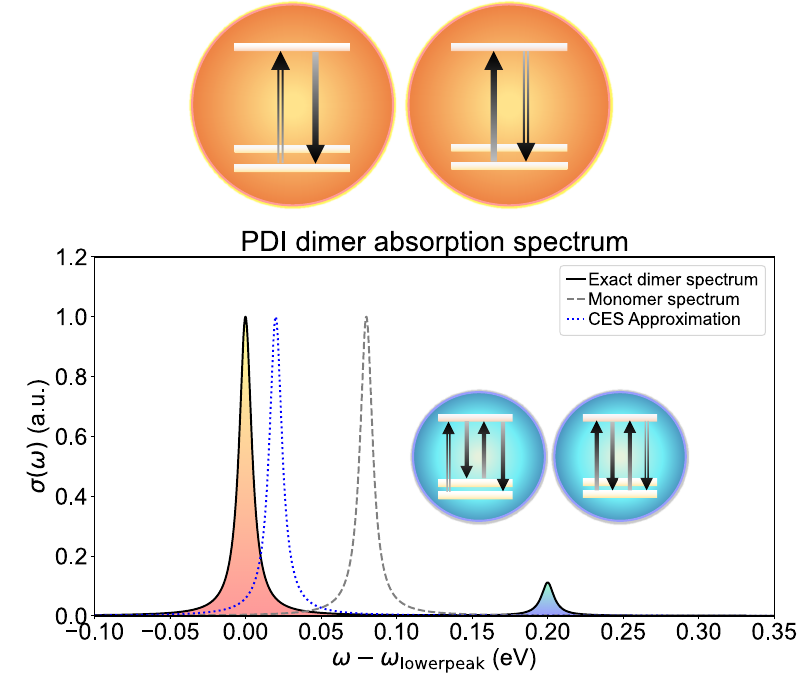}
    \caption{Absorption spectrum of a model PDI dimer illustrating the microscopic origin of Raman sidebands. The calculation includes a minimal vibronic manifold comprising a single excited-state vibrational level and two ground-state vibrational levels, sufficient to capture the essential physics of sideband formation. The orange peak corresponds to the feature captured by the DDA/CPA/CES approximation, which physically represents Rayleigh-type pathways, as schematically depicted through the ladder diagrams~\cite{tokmakoff_nonlinear_notes} in the orange figure. In contrast, the blue peak arises from vibronic sidebands neglected in the DDA/CPA/CES approximation; these features originate from Raman-like processes, as illustrated through the ladder diagrams~\cite{tokmakoff_nonlinear_notes} in the blue schematic. The parameters used in the computation are mentioned in the supplementary information section IIB}
\label{fig: dimer}
\end{figure}

\noindent We consider a PDI--PDI dimer~\cite{pochas2013contrasting}, whose exact absorption spectrum is shown in Fig.~\ref{fig: dimer}, alongside that of the PDI monomer (see supplementary information section IIB). For analytical clarity, we restrict the model to two vibrational levels on the ground-state potential energy surface and a single vibronic level on the excited-state surface, with Franck--Condon overlaps set to unity. The physical conclusions are independent of this truncation and extend straightforwardly to an arbitrary number of vibronic levels.\\

\noindent Within this three-level description, the exact dimer response is fully captured by retaining only the zeroth- and first-order contributions shown in Fig.~\ref{fig: dimer}. Higher-order manifolds do not arise in this case, since at most one monomer in the dimer can host a vibrational excitation in the electronic ground state.\\

\noindent The dominant dimer exact absorption peak in Fig.~\ref{fig: dimer} is red-shifted relative to the monomer resonance, consistent with the characteristic spectral signature of J-aggregate–type coupling. This primary feature is largely captured within the CPA framework (equivalently, the zeroth-order approximation), up to a small residual shift which we discuss below. \\

\noindent Beyond the dominant peak, the dimer spectrum exhibits additional features displaced from the CPA prediction. Notably, these shifts coincide with the ground-state vibrational frequencies of the PDI molecule. Moreover, this additional vibrational feature couples with the CPA peak, resulting in a further red shift of the orange peak in Fig.~\ref{fig: dimer} relative to the CPA prediction.\\

\noindent Thus, this result demonstrates that the corrections to the CPA originate from Raman-type processes that generate vibrational excitations in the electronic ground state. In a relevant homo-dimer, these processes manifest as sidebands to the CPA peaks, thereby encoding the Raman vibrational information in their linear response.

\section{conclusion}
\label{sec: conclusion}

\noindent In this work, we present a unified theoretical framework that weaves together vibronic structure, symmetry constraints, and an intrinsic separation of dynamical timescales, revealing that the linear absorption of all-to-all coupled aggregates is governed by two universal classes of photophysical pathways: Rayleigh-type and Raman-type scattering processes. In the thermodynamic limit, this framework yields a strikingly simple absorption spectrum, in which collective aggregate features are inherited from the monomer response through a simple, analytically controlled renormalization by the coupling parameter, $J$. Thus, in this limit, DDA/CPA/CES emerges not as uncontrolled approximations but as exact descriptions of infinitely large all-to-all coupled aggregates, providing a physically transparent reference theory for interpreting CPA-based spectra across aggregates of arbitrary size and connectivity. \\

\noindent The underlying mean-field approximation done in DDA/CPA/CES suppresses Green’s-function pathways involving ground-state vibrational excitations. By systematically reinstating the neglected terms at this mean-field level, we derive corrections beyond the CPA and explicitly compute them for the simplest nontrivial case of a molecular dimer. When applied to a PDI–PDI dimer, our analysis uncovers Raman-type vibronic features that are absent in CPA. These appear as additional absorption peaks displaced from the CPA-captured peaks by the ground-state vibrational frequencies, thereby revealing otherwise hidden molecular information encoded directly in linear absorption spectra. \\

\noindent More generally, our results establish a controlled microscopic limit in which classical-optics descriptions of aggregate spectra become exact, while simultaneously identifying the leading quantum-optical processes that arise beyond this limit. In doing so, the present framework places DDA, CPA and CES on a firmer microscopic foundation, clarifies both the origin of their success and the nature of their breakdown, and demonstrates that linear spectra of molecular aggregates can encode vibrational and Raman-type signatures that are absent at the mean-field level. Taken together, these results sharpen the structure-spectra relationship in molecular aggregates and provide a unified theoretical framework for determining when classical optical descriptions are sufficient and when explicitly quantum vibronic effects must be retained.\\

\begin{acknowledgments}

We are pleased to dedicate this work to the celebration of Jianshu Cao's legacy in chemical physics, which is characterized by great analytical elegance and physical insight. His pioneering works on the CPA within the context of molecular aggregates set the stage for the research done in this article. This work was supported with a Camille Dreyfus Teacher-Scholar Award. S.R.-C. thanks Juan B. Pérez-Sánchez and Kai Schwennicke for useful discussions.

\end{acknowledgments}

\section*{AUTHOR DECLARATIONS}

\subsection*{Conflict of Interest}
The authors have no conflicts to disclose.

\subsection*{Data Availability}
Data sharing is not applicable to this article as no new data were created or analyzed in this study.

\bibliography{references}

\clearpage
\onecolumngrid

\setcounter{section}{0}
\setcounter{equation}{0}
\setcounter{figure}{0}
\setcounter{table}{0}

\begin{center}
\textbf{\large Supplementary Information: Permutationally symmetric molecular aggregates}\\[6pt]
Sricharan Raghavan-Chitra, Arghadip Koner, Joel Yuen-Zhou\\
\textit{Department of Chemistry and Biochemistry, University of California San Diego, La Jolla, California 92093}
\end{center}
\vspace{1cm}

\section{Exact linear response of permutationally symmetric molecular aggregates - Continued Fraction Expression}

\noindent Let us consider the block-tridiagonal Hamiltonian given in Eq.~12 of the main text:

\begin{equation}
H^{(1)}=
\begin{pmatrix}
H_{e,0} & v_{0} & 0 & \cdots & 0 \\
v_{0}^{\dagger} & H_{e,1} & v_{1} & \ddots & \vdots \\
0 & v_{1}^{\dagger} & H_{e,2} & \ddots & 0 \\
\vdots & \ddots & \ddots & \ddots & v_{N-1} \\
0 & \cdots & 0 & v_{N-1}^{\dagger} & H_{e,N}
\end{pmatrix}.
\label{eq: si_block_tridiagonal}
\end{equation}

\noindent The linear response of a system of all-to-all coupled aggregate system made up of $N+1$ monomers is given by 
\begin{equation}
    \sigma(\omega)\propto-\Im\langle\hat{\mu}\hat{G}(\omega)\hat{\mu}\rangle,
\end{equation}

\noindent where for a zero-temperature initial state, the expectation value is taken with respect to
\begin{equation}
\left|N+1,0,\dots,0;\,0,\dots,0\right\rangle,
\end{equation}
for which all $N+1$ monomers occupy the global ground state. The collective transition dipole operator is
\begin{equation}
\mu=\sum_i \mu_i,
\end{equation}
namely, the symmetric sum of the single-monomer transition dipole operators. In the second-quantized notation (see Eq. 8 in main text), this operator can be written as
\begin{equation}
\mu = |\mu_M|\sum_{mn}^{M}\left(
\left\langle \psi_n^{(g)} \big| \psi_m^{(e)} \right\rangle b_n^\dagger B_m
+
\left\langle \psi_m^{(e)} \big| \psi_n^{(g)} \right\rangle b_n B_m^\dagger
\right),
\end{equation}
where $|\mu_M|$ denotes the magnitude of the transition dipole moment of an individual monomer. Thus, the expectation value entering the linear response becomes
\begin{equation}
\langle\mu G(\omega)\mu\rangle=(N+1)|\mu_{M}|^{2}\sum_{mn}^{M}\left\langle \psi_{0}^{(g)}\big|\psi_{n}^{(e)}\right\rangle \left\langle \psi_{m}^{(e)}\big|\psi_{0}^{(g)}\right\rangle \left\langle \alpha_{m}\right|G(\omega)\left|\alpha_{n}\right\rangle,
\label{eq: mu_G_mu_in_terms_of_alpha}
\end{equation}

\noindent where, $|\alpha_n\rangle \equiv |N,0,\dots,0;\,0,\dots,1_n,\dots,0\rangle$.

\noindent Here, the retarded Green's function corresponding to this Hamiltonian is

\begin{equation}
G(\omega) = \left[\omega - H^{(1)} + i0^{+}\right]^{-1}.
\end{equation}

\noindent We now employ the Schur complement to evaluate the expectation value in Eq.~\ref{eq: mu_G_mu_in_terms_of_alpha}, following a procedure similar to that used in the polariton literature (see Sec.~3B of the SI in Ref.~\cite{koner2025hidden}). For a block matrix
\begin{equation}
M=
\begin{pmatrix}
A & B\\
C & D
\end{pmatrix},
\end{equation}
provided the required inverses exist, its inverse can be written as
\begin{equation}
M^{-1}
=
\begin{pmatrix}
(M/D)^{-1} & -A^{-1}B(M/A)^{-1} \\
-D^{-1}C(M/D)^{-1} & (M/A)^{-1}
\end{pmatrix},
\end{equation}
where the Schur complements of $D$ and $A$ are defined, respectively, by
\begin{equation}
M/D := A - B D^{-1} C,
\qquad
M/A := D - C A^{-1} B.
\end{equation}

\noindent In what follows, we will primarily make use of the Schur complement of $D$.

\noindent\noindent We are interested in matrix elements of the Green’s function of the form
$\langle \alpha_n | G(\omega) | \alpha_m \rangle$ (see Eqs.~\ref{eq: mu_G_mu_in_terms_of_alpha}). This corresponds to evaluating the block of $(\omega - H^{(1)} + i\epsilon)^{-1}$ projected onto the first sector associated with $H_{e,0}$ in Eq.~\ref{eq: si_block_tridiagonal}.

\noindent Using the Schur complement, the projected resolvent can be expressed as
\begin{equation}
\langle \mu G(\omega)\mu \rangle
=
\left\langle
\mu
\left[
\omega - H_{e,0} + i\Gamma_{0}
- v_{0}
\left(\omega - R_{1} + i\Gamma_{1}\right)^{-1}
v_{0}^{\dagger}
\right]^{-1}
\mu
\right\rangle.
\end{equation}

\noindent Here, linewidths are incorporated phenomenologically via
$\Gamma_0 = \gamma/2$ and $\Gamma_k = (\gamma + \gamma_v)/2$ for $k \geq 1$, where $\gamma$ and $\gamma_v$ denote the homogeneous decay rates of the electronic and vibrational excitations, respectively.

\noindent Here,

\begin{equation}
R_{k} = Q_{k} H^{(1)} Q_{k},
\qquad
Q_{k} = \sum_{j=k}^{N} P_{j},
\end{equation}

\noindent where $P_j$ is the projector of the Hilbert subspace corresponding to $H_{e,j}$.

\noindent This procedure can be repeated iteratively to obtain

\begin{equation}
\langle \mu G(\omega)\mu \rangle
=
\left\langle
\mu
\left[
\omega - H_{e,0} + i\frac{\gamma}{2}
- v_{0}
\frac{1}{
\omega - H_{e,1} + i\frac{\gamma+\gamma_{v}}{2}
- v_{1}\frac{1}{\omega - R_{2} + i\Gamma_{2}}v_{1}^{\dagger}
}
v_{0}^{\dagger}
\right]^{-1}
\mu
\right\rangle.
\end{equation}

\noindent Finally, after repeating the iteration $N$ times, we obtain the expression given in the main text (Eq.~13):

\begin{equation}
\sigma(\omega)\propto-\Im\langle\mu G(\omega)\mu\rangle=-\Im\left\langle \mu\left[\omega-H_{e,0}+i\frac{\gamma}{2}-v_{0}\frac{1}{\omega-H_{e,1}+i\frac{\gamma+\gamma_{v}}{2}-v_{1}\frac{1}{\ddots-v_{N-1}\frac{1}{\omega-H_{e,N}+i\frac{\gamma+\gamma_{v}}{2}}v_{N-1}^{\dagger}}v_{1}^{\dagger}}v_{0}^{\dagger}\right]^{-1}\mu\right\rangle 
\label{eq: operator_continued_fraction}
\end{equation}

\section{Analytical example: $\Lambda$ system}

\subsection{Explicit block diagonal matrix and exact linear response}

\noindent Let us consider a $\Lambda$-type system in a truncated basis comprising two vibrational levels in the electronic ground-state potential energy surface and one vibrational level in the electronically excited potential energy surface. We further assume that the Franck--Condon overlaps between these states are unity. This minimal basis is sufficient to capture the class of photophysical processes that contribute to the linear response of the all-to-all coupled aggregate system. Accordingly, the expectation value entering the linear response (see Eq.~\ref{eq: mu_G_mu_in_terms_of_alpha}) reduces to
\begin{equation}
\langle\mu G(\omega)\mu\rangle=(N+1) |\mu_M |^2\left\langle N,0;1\right|G(\omega)\left|N,0;1\right\rangle .
\label{eq: three_level_system_exp_value}
\end{equation}

\noindent Further, the Hamiltonian in (Eq. 10 of main text) reduces to
\begin{equation}
\begin{aligned}
H=\omega_{g,1}b_{1}^{\dagger}b_{1}+\omega_{e,0}B_{0}^{\dagger}B_{0}+J(b_{0}^{\dagger}B_{0}^{\dagger}B_{0}b_{0}+b_{1}^{\dagger}B_{0}^{\dagger}B_{0}b_{1}),
\end{aligned}
\label{eq: bosonic_all_to_all_three_level_system}
\end{equation}

\noindent where, we have considered $\omega_{g,0} = 0$ for simplicity. 
Thus, the matrices $H_{e,0}$, $H_{e,1}$, $H_{e,2}$, $v_{0}$, and $v_{1}$ in Eq.~\ref{eq: si_block_tridiagonal} may be written explicitly for pedagogical purposes as

\begin{equation}
H_{e,0}=(\omega_{e,0}+NJ)\,|N,0;1\rangle\langle N,0;1|,
\qquad
H_{e,1}=(\omega_{g,1}+\omega_{e,0}+NJ)\,|N-1,1;1\rangle\langle N-1,1;1|
\end{equation}

\begin{equation}
H_{e,2}=(2\omega_{g,1}+\omega_{e,0}+NJ)\,|N-2,2;1\rangle\langle N-2,2;1|,
\qquad
v_{0}=\sqrt{N}J\,|N,0;1\rangle\langle N-1,1;1|
\end{equation}

\begin{equation}
v_{1}=\sqrt{2}\sqrt{N-1}J\,|N-1,1;1\rangle\langle N-2,2;1|,
\qquad
v_{2}=\sqrt{3}\sqrt{N-2}J\,|N-2,2;1\rangle\langle N-3,3;1|
\end{equation}\\

\noindent Thus, the overall matrix appearing in Eq.~\ref{eq: si_block_tridiagonal} becomes
\begin{equation}
H=
\begin{pmatrix}
\omega_{e,0}+NJ & \sqrt{N}J & 0 & \cdots & 0\\
\sqrt{N}J & \omega_{g,1}+\omega_{e,0}+NJ & \sqrt{2}\sqrt{N-1}J & \ddots & \vdots\\
0 & \sqrt{2}\sqrt{N-1}J & 2\omega_{g,1}+\omega_{e,0}+NJ & \ddots & 0\\
\vdots & \ddots & \ddots & \ddots & \sqrt{N}J\\
0 & \cdots & 0 & \sqrt{N}J& N\omega_{g,1}+\omega_{e,0}+NJ
\end{pmatrix}.
\end{equation}

\noindent Using standard continued-fraction techniques familiar from the polariton literature, one obtains
\begin{equation}
\left\langle N,0;1\right|G(\omega)\left|N,0;1\right\rangle =\frac{1}{\omega-(\omega_{e,0}+NJ)+i\frac{\gamma}{2}-\frac{NJ^{2}}{\omega-(\omega_{g,1}+\omega_{e,0}+NJ)+i\frac{\gamma+\gamma_{v}}{2}-\frac{\ddots}{\omega-([N-1]\omega_{g,1}+\omega_{e,0}+NJ)+i\frac{\gamma+\gamma_{v}}{2}-\frac{NJ^{2}}{\omega-(N\omega_{g,1}+\omega_{e,0}+NJ)+i\frac{\gamma+\gamma_{v}}{2}}}}}
\label{eq: si_continued_fraction_three_level_system}
\end{equation}

\subsection{Dimer spectra}

\noindent In this section, we compute the exact absorption spectrum of a homodimer using the expression derived (see Eq.~\ref{eq: si_continued_fraction_three_level_system}) for an all-to-all coupled molecular aggregate comprising an arbitrary number of monomers, within the truncated basis consisting of two vibrational levels in the electronic ground state and one vibrational level in the electronically excited state. Since, our model considers aggregates made up of $N+1$ monomers, a homodimer would correspond to the case of $N=1$.

\noindent Thus, for $N=1$, Eq.~\ref{eq: si_continued_fraction_three_level_system} simplifies to
\begin{equation}
\left\langle N,0;1\right|G(\omega)\left|N,0;1\right\rangle
=
\frac{1}{
\omega-(\omega_{e,0}+NJ)+i\frac{\gamma}{2}
-\frac{NJ^{2}}{
\omega-(\omega_{g,1}+\omega_{e,0}+NJ)+i\frac{\gamma+\gamma_{v}}{2}
}
}.
\end{equation}

\noindent Using the parameters for a PDI--PDI dimer~\cite{pochas2013contrasting} listed in Table~\ref{Tab: dimer_parameters}, the absorption spectrum,
\begin{equation}
\sigma(\omega)\propto -\Im \left\langle N,0;1\right|G(\omega)\left|N,0;1\right\rangle,
\end{equation}
is plotted in Fig.~4 of the main text.

\section{zeroth order response of all-to-all coupled molecular aggregate system}

\noindent We now compute the linear response of the all-to-all coupled aggregate (see Eq.~\ref{eq: operator_continued_fraction}) in the limit $N\to\infty$ with $NJ$ held fixed. In this limit, the response is determined by the zeroth-order block of the Hamiltonian,
\begin{equation}
G_{\mathrm{zeroth}}(\omega)=\frac{1}{\omega-H_{e,0}+i\frac{\gamma}{2}}.
\end{equation}

\noindent For analytical simplicity, we again restrict attention to the truncated basis consisting of two vibrational levels in the electronic ground-state potential energy surface and one vibrational level in the electronically excited potential energy surface, and we assume unit Franck--Condon overlaps between these states. On this basis, the expectation value that enters the linear response is $\left\langle N,0;1 \right|G(\omega)\left| N,0;1 \right\rangle$ (see Eq.~\ref{eq: three_level_system_exp_value}). Moreover,
\begin{equation}
H_{e,0}=(\omega_{e,0}+NJ)\left|N,0;1\right\rangle\left\langle N,0;1\right|,
\end{equation}
and let,
\begin{equation}
H_M=\omega_{e,0}\left|N,0;1\right\rangle\left\langle N,0;1\right|.
\end{equation}
It then follows that
\begin{align}
\left\langle N,0;1\right|G_{\mathrm{zeroth}}(\omega)\left|N,0;1\right\rangle
&=
\left\langle N,0;1\right|
\frac{1}{\omega-H_{e,0}+i\frac{\gamma}{2}}
\left|N,0;1\right\rangle \\
&=
\frac{1}{\omega-(\omega_{e,0}+NJ)+i\frac{\gamma}{2}} \\
&=
\frac{1}{\omega-\omega_{e,0}+i\frac{\gamma}{2}}
\frac{1}{1-\frac{NJ}{\omega-\omega_{e,0}+i\frac{\gamma}{2}}} \\
&=
\frac{\left\langle N,0;1\right|g(\omega)\left|N,0;1\right\rangle}
{1-NJ\left\langle N,0;1\right|g(\omega)\left|N,0;1\right\rangle}, \label{eq: si_final_CES_derivation}
\end{align}
where we have defined,
\begin{equation}
g(\omega)=\frac{1}{\omega-H_M+i\epsilon},
\end{equation}
since the linear response of $N$ uncoupled monomers is 
\begin{equation}
    \sigma_{\text{uncoupled}}(\omega)\propto-\Im\langle\mu g(\omega)\mu\rangle.
\end{equation}

\noindent Hence, from Eq.~\ref{eq: si_final_CES_derivation},
\begin{equation}
\sigma_{\mathrm{zeroth}}(\omega)\propto -\Im \langle \mu G_{\mathrm{zeroth}}(\omega)\mu\rangle
=
-\Im
\frac{\langle \mu g(\omega)\mu\rangle}
{1-NJ\langle \mu g(\omega)\mu\rangle}.
\label{eq: SI_ces_cpa_equivalence}
\end{equation}

\noindent Notably, Eq.~\ref{eq: SI_ces_cpa_equivalence} remains unchanged irrespective of the basis truncation, and all the complexity of the monomer potential energy surfaces is embedded in $g(\omega)$.

\subsection{Surrogate system}

\noindent
The zeroth-order linear response derived in Eq.~(\ref{eq: SI_ces_cpa_equivalence}) is exact in the limit $N \to \infty$ while $NJ$ is kept constant. Further, this result coincides with expressions resulting from implementing the DDA/CPA/CES approximation~\cite{chenu2017construction, eisfeld2002j} on molecular homoaggregates with arbitrary geometric arrangement.\\

\noindent
Thus, the \textit{approximate} DDA/CPA/CES response of a real system becomes the exact response of a surrogate system. For example, the linear response of a system of nearest-neighbour–coupled linear molecular aggregates under the CES approximation~\cite{eisfeld2002j} is
\begin{equation}
\sigma_{\text{NN}}(\omega)\propto -\Im\frac{G_{e,0}}{1-2JG_{e,0}},
\label{Eq: nearest_neighbour_CPA}
\end{equation}
where $G_{e,0}$ is a monomer property that can be computed from the monomer response, $\sigma_{m}(\omega)\propto -\Im G_{e,0}$, and $2J$ is the sum of the couplings between a monomer and its neighbors.
\\

\noindent
The essence of the CES approximation is usually described mathematically, pointing to the underlying mean-field approximation. However, the physical relevance of this approximation is often not stated clearly. Here, we find that the CES-approximated response in Eq.~(\ref{Eq: nearest_neighbour_CPA}) can be understood as the exact response of a surrogate aggregate in the infinite-monomer limit of an all-to-all–coupled system,
\begin{equation}
\sigma_{\text{all-to-all}}(\omega)\propto -\Im\frac{G_{e,0}}{1-\tilde{N}\tilde{J}G_{e,0}},
\end{equation}
in which the coupling between any two monomers is rescaled such that $\tilde{N}\tilde{J}=2J$, where $\tilde{N}\tilde{J}$ is the collective coupling strength of the all-to-all–coupled aggregate. Thus, there is always a surrogate all-to-all coupled molecular aggregate where the DDA/CPA/CES approximation is exact.

\begin{table}[h]
\centering
\caption{Parameters used for the dimer absorption spectra~\cite{pochas2013contrasting} shown in Fig.~4 of the main text. All energies are given in eV.}

\begin{tabular}{|c|c|}
\hline
\textbf{Parameter} & \textbf{Value (eV)} \\
\hline\hline
$\omega_{e,0}$ & 2.3 \\
\hline
$\omega_{g,1}$ & 0.16 \\
\hline
$J$ & -0.06 \\
\hline
$\gamma$ & 0.01 \\
\hline
$\gamma_{v}$ & $10^{-5}$ \\
\hline
\end{tabular}
\label{Tab: dimer_parameters}

\end{table}

\end{document}